\documentclass[useAMS,usenatbib]{mn2e}
\usepackage{chngpage}
\usepackage{graphicx}
\usepackage{natbib}
\usepackage{url}
\usepackage{amsmath}
\usepackage{booktabs}
\usepackage{threeparttable}
\usepackage{cite}
\usepackage[usenames,dvipsnames]{color}
\usepackage{amsmath}
\usepackage{amssymb}

\title[A new FRB]
  {A new fast radio burst in the datasets containing the Lorimer burst}
  \author[Zhang et al.]
    {S.-B. Zhang$^{1,2,3,4}$, G. Hobbs$^{3}$, S. Dai$^{3}$, L. Toomey$^{3}$, L. Staveley-Smith$^{4,5}$,
    \newauthor C. J. Russell$^{6}$  and X.-F. Wu$^{1}$\\
  $^{1}$Purple Mountain Observatory, Chinese Academy of Sciences, Nanjing 210008, China\\
  $^{2}$University of Chinese Academy of Sciences, Beijing 100049, China\\
  $^{3}$CSIRO Astronomy and Space Science, Australia Telescope National Facility, Box 76, Epping, NSW 1710, Australia\\
  $^{4}$International Centre for Radio Astronomy Research, University of Western Australia, Crawley, WA 6009, Australia\\
  $^{5}$ARC Centre of Excellence for All Sky Astrophysics in 3 Dimensions (ASTRO 3D)\\
  $^{6}$CSIRO Scientific Computing Services, Australian Technology Park, Locked Bag 9013, Alexandria, NSW 1435, Australia\\}

\begin{document}

\maketitle

\begin{abstract}
We report the discovery of a new fast radio burst (FRB), FRB~010312, in archival data from a 1.4\,GHz survey of the Magellanic Clouds using the multibeam receiver on the Parkes 64\,m-diameter radio telescope. These data sets include the Lorimer burst (FRB~010724), which it pre-dates and which we also re-detect. The new burst has a much higher dispersion measure of 1187\,cm$^{-3}$pc. The burst is one of the broadest found to date, the second earliest FRB known, and the ninth FRB discovered with a dispersion measure larger than 1000\,cm$^{-3}\,$pc. Our discovery indicates that there are likely to be more burst events still to be found in the existing Parkes data archive.
\end{abstract}


\section{Introduction}
\label{sec:intro}

The first reported fast radio burst (FRB) was detected as a bright single pulse of millisecond duration using the Parkes 64\,m-diameter radio telescope~\citep{Lorimer07}. The FRB was discovered by reprocessing archival observations that had originally been observed in order to search for pulsars in the Magellanic clouds~\citep{Manchester06}. Those observations are available in the ATNF pulsar data archive~\citep{Hobbs11} and, in this paper, we report the discovery of a second FRB in the same data set.

Since the initial discovery, more than 60 FRBs have been published\footnote{The FRB online Catalogue is available from \url{http://www.frbcat.org}}~\citep{Petroff16}  and their high dispersion measure (DM), in excess of the expected contribution from the Milky Way, implies an extragalactic origin. Furthermore, the localization of one repeating FRB to a dwarf galaxy at redshift $z \thicksim 0.2$ ~\citep{Spitler16,Chatterjee17,Tendulkar17}, is consistent with the DM-estimated distance and, assuming that the repeating FRBs are similar in origin to those discovered by the Parkes telescope, confirms that FRB progenitors are at cosmic distances. The most successful telescopes for finding FRBs have been Parkes~\citep{Lorimer07,Keane12,Thornton13,Burke-Spolaor14,Ravi15,Petroff15,Keane16,Champion16,Ravi16,Petroff17,Bhandari18,Petroff19}, Molonglo~\citep{Caleb17}, ASKAP~\citep{Shannon18,Macquart18} and, most recently, CHIME~\citep{CHIME19a,CHIME19b}. The distribution of the DMs for the detected FRB population seems to vary between telescopes. This is explained by a combination of the telescope sensitivity and field-of-view with various models of FRB event rates~\citep{Shannon18}. In particular, the four FRBs with the highest DMs were discovered with the Parkes telescope (which, including the one presented here, has found seven out of the nine known FRBs with a DM larger than 1000\,cm$^{-3}$pc).

The progenitors of FRBs remain mysterious, although plenty of models have been proposed. According to their short duration ($\thicksim$ ms) and extremely high inferred brightness temperature ($\gtrsim 10^{35} $ K), a coherent emission process~\citep{Melrose16} has been invoked to explain the radiation. Such emission processes include magnetic reconnections of a neutron star's magnetosphere caused by the outflow from a super-massive black hole in the host galaxy~\citep{Zhang18} and supergiant pulses from extragalactic pulsars~\citep{Cordes16}. Other models involve merging neutron stars~\citep{Totani13,Wang16} or white dwarf stars~\citep{Kashiyama13}, collapse of super-massive neutron stars to black holes~\citep{Falcke14}, magnetar flares~\citep{Popov13}, neutron star-white dwarf binary accretion~\citep{Gu16} and even cosmic string collisions~\citep{Cai12}, collisions between neutron stars and asteroids~\citep{Huang16,Dai16} or charged black hole mergers~\citep{Zhang16,Liu16,Deng18}. However, most current models cannot yet be confirmed, nor ruled out, because of the insufficient sample of FRBs.         
  
In this letter, we present a new FRB (known as FRB~010312) detected during reprocessing of archival data from a 1.4\,GHz survey of the Magellanic Clouds.   This FRB is the second earliest FRB yet detected, one of the widest, and the ninth FRB discovered with DM larger than 1000\,cm$^{-3}\,$pc. In Section~\ref{sec:data}, we describe the details of the observations and data reduction. 
The properties of the burst and discussion of our detection are presented in Section~\ref{sec:result_dis}. We conclude in Section~\ref{sec:con}

\section{Observation and Data Reduction}
\label{sec:data}

\citet{Manchester06} carried out the survey of the large and small Magellanic Clouds with the primary goal of discovering pulsars. The survey was undertaken with the Parkes 64\,m-diameter radio telescope between May 2000 to August 2001 and led to the discovery of 14 pulsars.  Other known pulsars were redetected and the survey observations also included observations of bright known pulsars, such as PSR~J0437$-$4715, in order to confirm that the system was working correctly.  The observations for this survey have been archived in the Parkes data archive (\url{https://data.csiro.au}; see \citet{Hobbs11} for details) using the project code identifier P269. The survey used the 21\,cm multibeam receiver centred at $1374$\,MHz. The channelised and polarization-summed signals were one-bit sampled and recorded using an Analogue Filter Banks (AFB) system. The bandwidth, number of channels and sampling time  of these observations are 288\,MHz, 96 and 1\,ms, respectively. 

\citet{Lorimer07} searched for bright single pulses up to a DM of 500\,cm$^{-3}$pc and successfully identified a burst with a DM of 375\,cm$^{-3}$pc, which is now known to be the first detection of an FRB.  There are no published searches of the same data set to higher DMs.   We are carrying out a project to reprocess all the search-mode observations in the data archive to search for events such as FRBs. We started with the Magellanic Cloud survey primarily to confirm that we can re-detect the~\citet{Lorimer07} event.  The survey data contains approximately 6250\,h of on-sky integration time. This corresponds to  $\sim$\,267 deg$^2$\,h.  A recent Parkes survey with the same receiver system (the HTRU survey) had an FRB event rate of 1/144\,deg$^{-2}$h$^{-1}$~\citep{Champion16}. We therefore predicted that the Magellanic Cloud survey data should include around two FRBs.

We used the pulsar searching software package \emph{\sc presto}\footnote{\url{http://www.cv.nrao.edu/~sransom/presto/}}~\citep{Ransom01} and processed the data on CSIRO's high performance computer  facilities. Strong narrow-band and short-duration broadband radio frequency interference (RFI) were identified and marked using the \emph{\sc presto} routine \emph{\sc rfifind}. We used a 1\,s integration time for our Radio Frequency Interference (RFI) identification and the default cutoff to reject time-domain and frequency-domain interference in our pipeline. To avoid deleting possible bursts, we used the option \emph{\sc noclip} during all the processing. In preparation for de-dispersion, the \emph{\sc ddplan.py} algorithm was used to determine the DMs required for us to search. The DM range that we searched was $0$ to $5000\,\,$cm$^{-3}\,$pc and the number of DM trials was 440.
Data were then de-dispersed at each of the trial DMs using the \emph{\sc prepdata} routine, and RFI was removed based on the mask file. Single pulse candidates with a signal-to-noise ratio (S/N) larger than five were identified using the \emph{\sc single\_pulse\_search.py} routine for each de-dispersed time series\footnote{We used the command-line option {\tt -b} and boxcar filtering with filter widths of 1, 2, 3, 4, 6, 9, 14, 20, 30, 45, 70, 100, 150, 220 and 300 sample. The maximum search was equivalent to a width of 300\,ms. We use the definition of $\sigma$ as presented by \textsc{presto} for our S/N value.}. All of the several tens of thousands of candidates were ranked and plotted using the same method as~\citet{ZhangS18}. Those that were not clearly caused by RFI were visually inspected.

\section{Results and discussion}
\label{sec:result_dis}

\begin{table}
\caption{The properties of FRB~010312.}
\begin{center}
\begin{threeparttable}
\begin{tabular}{ll}
\hline
\hline
{\bf Observed Properties} \\
Event data UTC & 2001 March 12\\
Event time UTC, $\nu_{1.374 \rm {GHz}}$ & 11:06:47.98\\
Event time Local (AEDT), $\nu_{1.374 \rm {GHz}}$ & 22:06:47.98\\
Pointing R.A. (J2000) & 05:26:54.9 \\
Pointing Dec. (J2000) & $-$64:56:19.2 \\
Galactic longitude & 274.72$^{\circ}$ \\
Galactic latitude & $-$33.30$^{\circ}$ \\
Beam 7 full-width, half-maximum & 14.1$'$\\
DM (cm$^{-3}\,$pc) & 1187$\pm$14\\
Observed width (ms) & 24.3$\pm$1.3\\
S/N & 11\\
\hline
{\bf Inferred Properties}\\
Peak flux density (Jy) & 0.25\\
Fluence (Jy ms) & 6.1\\
DM$_{\rm MW, NE2001}$ (cm$^{-3}\,$pc)  & 51\\
DM$_{\rm MW, YMW16}$ (cm$^{-3}\,$pc)  & 55\\
DM$_{\rm MC, YMW16}$ (cm$^{-3}\,$pc)  & 12\\
Redshift$_{\rm YMW16}$, z & 1.4\tnote{a}\\
Distance$_{\rm YMW16}$ (Gpc) & 3.9\tnote{a}\\
\hline
\end{tabular}
   \begin{tablenotes}
        \footnotesize
        \item[a]The DM of host galaxy was assumed to be 100\,\,cm$^{-3}\,$pc and the calculation used the YMW16 model~\citep{Yao17}. 
     \end{tablenotes}
\end{threeparttable}
\end{center}
\label{table:properties}
\end{table}

In our processing, we re-detected the original FRB in the observation from 2001 July 24 at a DM of $375\,\,$cm$^{-3}\,$pc, in three beams (beams 6, 7 and 13) with S/N of 31, 16 and 26 respectively.  We therefore are confident in the quality of the archived data, our processing, visualisation and ranking methods.

\begin{figure}
\begin{center}
\begin{tabular}{ll}
\includegraphics[width=8cm,angle=0]{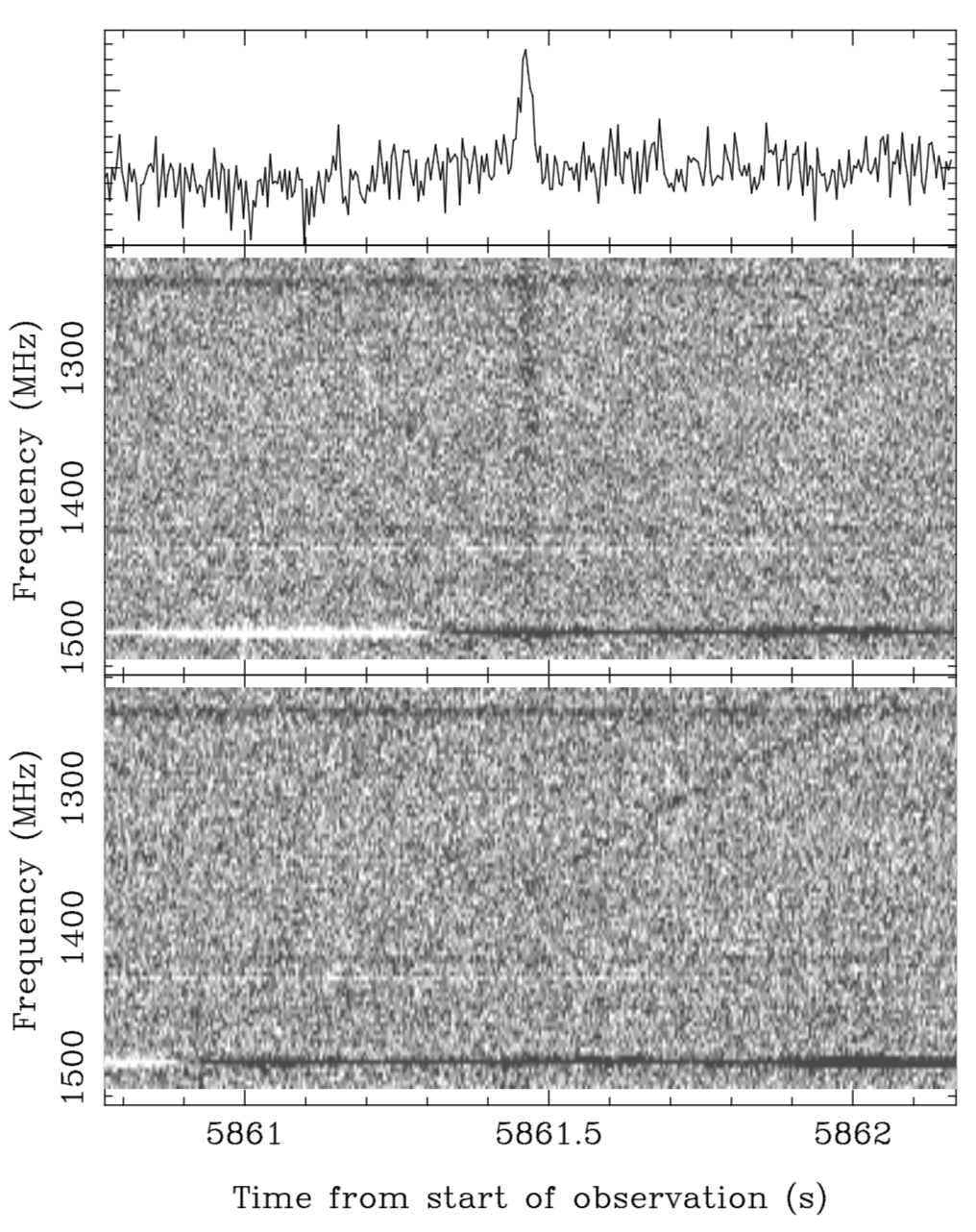} 
\end{tabular}
\caption{Frequency-time plane of the FRB~010312 without dispersion (bottom), with dispersion (central) and its integrated pulse profile (shown using an arbitrary flux scale) after being de-dispersed at the optimal DM value (top). The frequency resolution is 3\,MHz. The time axis shows the time since the start of the observation
(March 12, 2001 09:29:07 UTC) with a resolution of 4\,ms.} 
\label{figure:FRB}
\end{center}
\end{figure}

A new FRB with a DM of 1187$\pm$14\,\,cm$^{-3}\,$pc and S/N$=$11 was also detected in our search. Figure~\ref{figure:FRB} shows the burst in the frequency-time plane and its integrated pulse profile after being de-dispersed at the optimal DM value.   We can see that the signal is much stronger in the lower part of the observing band. This is similar to FRBs such as FRB~110214~\citep{Petroff19} and FRB~171019~\citep{Shannon18}. There are also two strong narrow-band RFI signatures close to the band edges, but they do not affect our ability to detect the FRB.   

The properties of FRB~010312 are presented in Table~\ref{table:properties}. Our FRB\footnote{Since the discovery of the first FRB searches have been undertaken in even earlier data sets. The earliest FRB yet detected is FRB~010125 again detected in archival Parkes data and our new FRB (FRB~010312)  is therefore the second-earliest known FRB.} 
was detected in only a single beam (Beam 7) of the multibeam receiver. The remaining 12 beams show no clear candidate, nor RFI around the time of the burst. We therefore cannot precisely provide the position of the source.  In Table~\ref{table:properties} we simply list the pointing position of the beam in right ascension and declination (and converted to Galactic coordinates). 

The burst has a width of 24.3$\pm$1.3\,ms at its 50\% power point, which is one of the largest values reported thus far (note that~\citet{Farah17} report an FRB with a width of 26\,ms). We have searched for changes in the pulse width as a function of observing frequency, but the S/N of our profile precludes any detailed analysis.  

 From the estimated S/N, DM and sky position we can infer various properties that are listed in the lower half of Table~\ref{table:properties}. The peak flux density was obtained from the single pulse radiometer equation~\citep{Cordes03} and the S/N measurement. The YMW16 electron density model~\citep{Yao17}, which assumes $H_0=67.3\,{\rm km}\,{\rm s}^{-1}\,{\rm Mpc}^{-1}$~\citep{Planck14} and the local intergalactic medium baryon density $n_{\rm IGM} = 0.16\,{\rm m}^{-3}$~\citep{Katz16}, indicates a large cosmic distance of 3.9\,Gpc with the assumption of the host galaxy DM $\sim$100\,\,cm$^{-3}\,$pc. 

The burst was only detected in one beam and
in order to check whether there is any evidence of repeated bursts from this FRB, we used 46.2\,h of archival observations whose pointing positions were within 0.5\,deg of the position of the beam in which FRB~010312 was detected. We searched a DM range from 1158 to 1178\,\,cm$^{-3}\,$pc with a DM step of 0.1\,\,cm$^{-3}\,$pc, but no new convincing candidates were detected. 

The results from processing the entire 6250\,h-long survey led to the detection of the original FRB, the new one as described above, and the expected single pulse detections from known pulsars. \citet{Foster18} presented a set of tests that can be applied to give confidence to any potential FRB discovery.  We have applied these tests to both FRB~010312 and the Lorimer burst. We obtain almost identical test results providing confidence that our new FRB is real.  However,  FRB~010312 is not bright enough to enable a detailed study of the spectral and scattering properties, although in common with many other FRBs, FRB~010312 is significantly weaker in part of the band. We found no further FRB candidate with S/N $\geqslant$ 8 although around a dozen candidates with S/N $\geqslant$ 7 were detected.  We will discuss these much weaker signals further when we have completed our larger-scale analysis of the entire data archive.

With the addition of FRB~010312, the FRB event rate of the Magellanic Clouds survey is 1/134\,deg$^{-2}$ h$^{-1}$, which is similar to the prediction of~\citet{Champion16} of 1/144 deg$^{-2}$ h$^{-1}$ from the Parkes HTRU survey, which was based on 10 FRB detections.  However, \citet{Petroff19} has recently found a further FRB in this same data set giving an updating event rate closer to our measurement of  1/131\,deg$^{-2}$\,h$^{-1}$.

\section{Conclusion}
\label{sec:con}
Since~\citet{Lorimer07} discovered the first FRB, new bursts continue to be discovered using archival data with different DM trials and new search algorithms~\citep[e.g.,][]{Burke-Spolaor14,Champion16,Petroff19}. There are many reasons why FRBs were missed in some searches, including the challenges of looking through large numbers of candidates and different methods for dealing with RFI. However, the most significant issue is simply the range of DMs searched.  Searching over a very large DM range is computationally expensive and can increase the false alarm rate, but, as presented here, can lead to new discoveries.  

We have reported here on a single FRB that occurred many years ago with a relatively simple observing system.  We therefore cannot do detailed follow-up, multi-wavelength observations, nor carry out an in-depth analysis of the scattering, scintillation or polarisation properties of the burst. However, our results do provide confidence in the event rate predictions.  The recent predictions indicated that two FRBs should be detectable in the Magellanic Cloud survey and so our results confirm that these predictions are reliable.  

Parkes is one of the most sensitive telescopes used to find FRBs and so can find relatively weak bursts (when compared, for example, to the ASKAP discoveries).  The FRB reported here is one of only a handful of FRBs with DMs higher than 1000\,cm$^{-3}$\,pc and has one of the largest isotropic energy of $4.3\times10^{33}$J\footnote{It is estimated using the definition in~\citet{Petroff16} with adopting $H_0=67.3\,{\rm km}\,{\rm s}^{-1}\,{\rm Mpc}^{-1}$, $\Omega_{\rm m}= 0.315$ and $\Omega_{\Lambda}= 0.685$~\citep{Planck14}}.   As we continue to search the Parkes archival data we expect to increase the sample size for these extremely high DM FRB events. Telescopes such as ASKAP and CHIME are now finding large numbers of FRBs~\citep{Shannon18,CHIME19a}.  However, the Parkes multibeam data sets provide an enormous repository of data obtained with a single telescope and receiver system.  Even though the data are all in a similar format, different research groups have processed their data sets in different ways. With the archive and supercomputing facilities available we now, for the first time, have the opportunity of processing all the data in a self-consistent manner. We are confident that we will discover new FRBs in this processing and obtain a reliable FRB event rate over decades of observations.

\section*{Acknowledgments}
The Parkes radio telescope is part of the Australia Telescope National Facility which is funded by the Australian Government for operation as a National Facility managed by CSIRO.  This paper includes archived data obtained through the CSIRO Data Access Portal (http://data.csiro.au). This work was supported by a China Scholarship Council (CSC) Joint PhD Training Program grant and the National Natural Science Foundation of China (Grant No. 11725314). Parts of this research were supported by the Australian Research Council Centre of Excellence for All Sky Astrophysics in 3 Dimensions (ASTRO 3D), through project number CE170100013.

\end{document}